\documentclass[conference,compsoc]{IEEEtran}
\ifCLASSOPTIONcompsoc
  \usepackage[nocompress]{cite}
\else
  \usepackage{cite}
\fi
\usepackage{amsmath,amssymb,amsfonts}
\usepackage{graphicx}
\usepackage{booktabs}
\usepackage{multirow}
\usepackage{array}
\usepackage{algorithm}
\usepackage{algpseudocode}
\usepackage{xcolor}
\usepackage{xspace}
\usepackage{url}
\usepackage{hyperref}
\hypersetup{
  colorlinks=false,
  pdfborder={0 0 0.7},
  linkbordercolor={0.75 0.15 0.15},
  citebordercolor={0.15 0.55 0.25},
  urlbordercolor={0.15 0.35 0.75},
  pdftitle={From Speech Tokens to Stolen Voices: Attacker-Chosen-Text Speaker Impersonation via Representation Inversion},
  pdfauthor={Ye Lu, Shen Wang, Runze Liu, Zhitao Ou, Haonan Huang, Yihan Yan, Fanghui Sun, Zhaoyang Zhang},
  pdfkeywords={speech privacy, model inversion attacks, speaker impersonation, speech language models}
}
\graphicspath{{imgs/}}

\newcommand{\stia}{\textsc{STIA}\xspace}
\newcommand{\result}[2]{\ensuremath{#1 \pm #2}}
\newcommand{\meandelta}[3]{\ensuremath{{#1}_{(#2\,#3)}}}

\begin{document}
\bstctlcite{stia-bib-control}

\title{From Speech Tokens to Stolen Voices: Attacker-Chosen-Text Speaker Impersonation via Representation Inversion}

\author{%
  \IEEEauthorblockN{%
    Ye Lu,
    Shen Wang,
    Runze Liu,
    Zhitao Ou,
    Haonan Huang,
    Yihan Yan,
    Fanghui Sun, and
    Zhaoyang Zhang}
  \IEEEauthorblockA{School of Cyberspace Science, Harbin Institute of Technology}
  \IEEEauthorblockA{Corresponding author: Zhaoyang Zhang (zhaoyang\_zhang@hit.edu.cn)}
}

\maketitle

\begin{abstract}
Speech interaction systems have gradually evolved from traditional cascaded
systems that use text as intermediate information to end-to-end speech language
models that directly process speech representations. In some applications, the
speech frontend runs on the user's device and sends speech representations to a
remote server for inference, while the original audio remains local. However,
these representations may still contain speaker identity information. If an
attacker obtains them, it may use this information to imitate the user's voice,
threatening their voice privacy. To investigate this risk, we propose \stia, a
Speech Token Inversion Attack that generates speech with attacker-specified text
in a victim's voice using a single short leaked representation, without access
to the original audio or its transcript. Specifically, \stia converts discrete,
continuous, or mixed speech representations into conditioning sequences and
trains a conditional flow-matching speech generation model using pairs of
different utterances from the same speaker in public speech datasets. The model
combines speaker characteristics contained in the leaked representation with the
specified text to generate speech in the victim's voice. Experiments show that,
with a Moshi representation leak corresponding to 3 seconds of speech, \stia
achieves an average CAM++ speaker cosine similarity of 0.78 and a target-text
word error rate of 3.80\%. These results show that speech frontend representations
reveal reusable speaker identity information and highlight the need to consider
privacy risks when designing speech frontend representation methods.
\end{abstract}


\section{Introduction}
\label{sec:introduction}

End-to-end speech language models increasingly use speech-native
representations during inference, rather than relying on a cascaded pipeline
that uses text as the intermediate modality between automatic speech
recognition (ASR), a large language model (LLM), and text-to-speech (TTS)
synthesis. These models typically use a speech frontend to convert audio into
acoustic tokens, continuous features, or both, preserving emotional cues,
intonation, and other prosodic information that text alone cannot fully
convey~\cite{defossez2024moshi,kimiteam2025kimiaudio,xu2025qwen3omni}.
In a split deployment, user audio can be converted locally into these
representations, which are then sent to a remote server for
processing, while raw waveforms remain on the user's device. This deployment
may create the misleading impression that transmitting frontend representations
rather than raw waveforms is sufficient to protect speaker privacy.

However, prior work has shown that frontend encoding does not remove all
speaker information. SpInv recovers speaker embeddings, which are vector
representations used to compare speaker identities, from speech-language-model
frontend representations~\cite{lu2026spinv}. Its results establish that identity
information can remain exposed through frontend representations even when the
original waveform is unavailable. Keeping raw audio local is therefore
insufficient, by itself, to protect speaker privacy.

Speaker-embedding recovery leaves an unresolved security question: can an attacker reuse
the identity information retained in frontend representations to generate new
content in the victim's voice? Embedding recovery alone
does not establish whether a leaked representation supports speech generation
that both preserves the speaker's vocal characteristics and accurately conveys
chosen text. Existing voice-cloning methods can synthesize speech for new text
from short reference recordings, but typically rely on access to the speaker's
audio~\cite{jia2018transfer,casanova2022yourtts,wang2023valle}. We therefore
investigate whether an attacker with only fragmented frontend representations
of the victim's speech, and no original audio, can generate speech that
preserves the victim's vocal characteristics and conveys attacker-chosen text
unrelated to the source utterance.

\begin{figure*}[!t]
  \centering
  \includegraphics[width=0.98\textwidth]{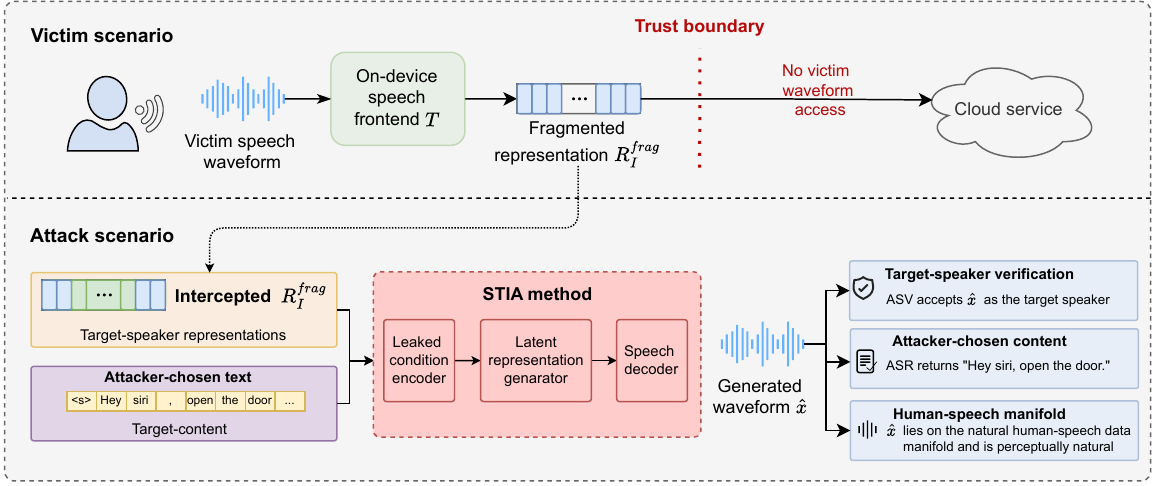}
  \caption{Split-deployment setting and \stia attack overview. The attacker
  combines leaked frontend representations with chosen text to generate speech;
  the boxes on the right state the attack objectives.}
  \label{fig:attack-scenario}
\end{figure*}

To investigate speaker impersonation from frontend representations, we propose
\stia, a Speech Token Inversion Attack.
Figure~\ref{fig:attack-scenario} places the attack in the split deployment we
consider: the victim's device converts speech into frontend representations
for a remote server, while retaining the raw waveform locally. We assume that
the attacker can read only the frontend representation of a segment of one
victim utterance, rather than the full-utterance representation. Neither the
victim's waveform nor the source transcript is available. \stia combines the leaked
fragmented representation with independently chosen text to generate a new
speech waveform.
The objectives on the right describe the intended impersonation capability:
the generated speech should preserve the victim's vocal identity, convey the
chosen text, and sound natural.

The central idea behind \stia is to learn to reuse vocal cues retained in
fragmented frontend representations to generate speech with new textual
content. During training, we use a public auxiliary speech dataset to learn a
conditional flow-matching model in a frozen audio codec's continuous latent
space. Each training example pairs a source representation fragment with
target text and speech from a different utterance by the same speaker.
Specifically, a leak conditioning encoder maps discrete, continuous, or mixed
representation fragments obtained through black-box frontend queries into
conditioning sequences. A trainable text projection maps the outputs of a
frozen multilingual text encoder to text-condition sequences of the same width.
Both conditions feed a Transformer shared between denoising and decoding modes.
The network learns a transformation from noise to target speech latents through
conditional flow matching, while codec-aware decoding supervision aligns
predicted latents with the target speech's discrete
codes~\cite{lipman2023flow,hu2026elf}. At attack time, the trained model
conditions on the victim's leaked representation fragment and attacker-chosen
text to iteratively generate speech latents from Gaussian noise. A final
latent decoding step is followed by quantization and waveform synthesis through
the frozen codec to produce the generated audio. No victim-specific training
or adaptation is required.

For evaluation, we select the frontends of three representative speech language
models (Moshi, Kimi-Audio, and Qwen3-Omni) covering discrete, mixed, and
continuous representations, respectively. We use speakers unseen during attack
training and assess vocal similarity against genuine recordings distinct from
the leakage source. We construct
SpInv+TTS as a two-stage baseline that recovers speaker embeddings before
synthesizing the chosen text with SpeechT5~\cite{ao2022speecht5}. Across all
three frontends, \stia achieves higher mean speaker similarity under both
CAM++ and ECAPA and lower mean target-text word error rate (WER), while mean
UTMOS scores for predicted naturalness remain similar to those of the baseline.
For a Moshi representation leak corresponding to 3 seconds of speech,
\stia obtains a mean CAM++ cosine similarity of 0.78 and a mean WER of
3.80\%. These results show that, under the evaluated conditions, fragmented
frontend representations can be used to generate speech that retains a victim's
vocal characteristics and conveys new, attacker-chosen content, creating risks
of voice impersonation and leakage of private speaker information.

Our contributions can be summarized as follows:
\begin{itemize}
  \item To the best of our knowledge, we are the first to study arbitrary-text
  speaker impersonation from leaked frontend representations of end-to-end
  speech language models. The only victim-specific input required is a short
  leaked representation fragment. The attacker can choose arbitrary target text
  independently of the source utterance, without access to the victim's waveform
  or source transcript.
  \item We propose \stia, which generates speech directly from leaked frontend
  representation fragments and attacker-chosen text. It accommodates discrete,
  continuous, and mixed representations, learns reusable vocal cues through
  same-speaker, cross-utterance training on public auxiliary speech, and combines
  conditional flow matching with codec-aware decoding supervision for synthesis,
  without victim-specific training data or a separate speaker-embedding recovery
  stage.
  \item We conduct extensive and detailed experimental evaluations
  on the frontends of three representative target models. The results show that \stia effectively exploits representation
  fragments to generate speech that
  preserves victims' vocal characteristics and conveys chosen text, outperforming
  the SpInv+TTS baseline method on both speaker-similarity metrics and mean
  target-text WER. We further examine the effects of leakage duration,
  auxiliary-data volume, model scale, and leakage and text conditioning, and
  test McAdams anonymization against a fixed attacker.
\end{itemize}

\section{Related Work}
\label{sec:related-work}

\subsection{Speech Representations and Speech--Language Interfaces}

Speech interfaces have progressed from task-specific acoustic features to
self-supervised representations and, subsequently, to discrete or continuous
interfaces consumed by speech language models. The wav2vec~2.0 contrastive
objective uses quantized latent targets, whereas HuBERT learns by masked
prediction of clustered hidden units; both provide
reusable frontends~\cite{baevski2020wav2vec2,hsu2021hubert}.  Recent systems
connect such representations to generative language models: SpeechGPT uses
discrete speech units for cross-modal conversation, Spectron directly models
spectrograms for spoken question answering and continuation, and Moshi supports
streaming full-duplex speech--text interaction~\cite{zhang2023speechgpt,nachmani2023spectron,defossez2024moshi}.
Such systems may expose token sequences or hidden states at module boundaries,
caches, or logs. We study the security consequence of this exposure: whether
a finite representation fragment retains speaker information reusable for
generation without access to the source waveform.

\subsection{Model Inversion and Privacy of Intermediate Representations}

Early model inversion examined how an adversary could recover training
inputs or sensitive features through a model interface. Work by
Fredrikson et al. exploited confidence outputs, while Zhang et al. later
introduced generative priors for inversion of deep
networks~\cite{fredrikson2015model,zhang2020secretrevealer}.  Recent studies show that
intermediate representations are also at risk: text embeddings can be inverted
to recover source text, including in multilingual
settings~\cite{morris2023text,chen2024multilingual}, while VoxGuard and HearSay broaden
systematic privacy evaluation for speech and audio language
models~\cite{tsaprazlis2026voxguard,wang2026hearsay}.

For vocal identity, Pizzi et al. reconstruct audio and intermediate voice
features from automatic speaker-recognition models and investigate
impersonation~\cite{pizzi2022inversion}. SpInv targets speech-language-model
frontend representations, inverting them into embeddings in an
attacker-specified speaker-encoder space~\cite{lu2026spinv}. This objective
recovers identity encodings rather than a generative representation that retains
all timbral and prosodic detail, potentially limiting voice fidelity when
combined with TTS. We adapt SpInv to this task by retraining it in the embedding
space required by SpeechT5 to construct a two-stage baseline; \stia instead
generates speech directly from leaked representations and target text.

\subsection{Speaker Recognition and Voice Anonymization}

Speaker recognition evolved from generative identity models to compact
embeddings and deep discriminative frontends.  Classic GMM--UBM systems used
adapted speaker models, i-vectors compressed speaker and channel variability
into a low-dimensional total-variability space, and x-vectors subsequently
learned utterance-level embeddings with deep
networks~\cite{reynolds2000speaker,dehak2011frontend,snyder2018xvectors}.  Recent
architectures such as ECAPA-TDNN and CAM++ strengthen temporal aggregation and
channel or context modeling for speaker
verification~\cite{desplanques2020ecapa,wang2023campp}.  In parallel, voice anonymization
seeks to suppress identity while retaining linguistic content: VoicePrivacy
established common threat models and evaluation protocols, with approaches
including learned pseudo-speaker mappings and signal transformations based on
the McAdams coefficient~\cite{tomashenko2022voiceprivacy,patino2021mcanonymization}.
These defenses primarily target identification or linkability; we test whether
representations exposed after attacker-unknown McAdams anonymization still
condition cross-content synthesis by the fixed \stia model.

\subsection{Zero-Shot Voice Cloning and Audio Deepfakes}

Few-shot multispeaker synthesis moved voice cloning from per-speaker training
to inferring a speaker condition from a short reference recording.  Classic
work by Arik et al. compared few-shot speaker adaptation and speaker encoding,
while Jia et al. transferred speaker-verification embeddings to multispeaker
text-to-speech synthesis~\cite{arik2018cloning,jia2018transfer}.  More recently,
YourTTS enabled zero-shot multispeaker and cross-lingual synthesis, and VALL-E
cast zero-shot TTS as neural audio-codec language
modeling~\cite{casanova2022yourtts,wang2023valle}.

Recent flow-matching methods further advance zero-shot speech generation.
Voicebox supports multilingual generation through text-guided speech infilling;
E2 TTS simplifies non-autoregressive synthesis, and F5-TTS improves
flow-matching TTS robustness through refined text modeling and
sampling~\cite{le2023voicebox,eskimez2024e2tts,chen2024f5tts}. These methods
typically derive voice conditions from reference audio and do not natively
accept arbitrary leaked representations from different black-box frontends.
\stia studies how to use such representations directly, without the victim's
waveform or source transcript. ASVspoof addresses the complementary detection
problem of distinguishing genuine speech from synthetic, converted, or replayed
speech~\cite{todisco2019asvspoof}.

\section{Threat Model and Motivation}
\label{sec:problem}

\subsection{Threat Model}

\subsubsection{Target model}
We consider a hypothetical split deployment of an open-source speech language
model: the user device runs the speech frontend $T$, and the remote service
runs the speech backend $G$. This setting keeps raw waveforms local instead of
uploading audio directly, allowing us to study the privacy risk of exposing
frontend representations.
Given user speech $x$, the frontend produces $R=T(x)$.  The backend combines
this representation with textual, system, or session context $p$ available to
the service.  Denoting the complete model by $M$, we have
\begin{equation}
 M(x,p)=G(R,p), \qquad R=T(x).
 \label{eq:split-model}
\end{equation}
Thus, the client must make $R$ available to the service, while the source
waveform $x$ need not leave the device.

We write the frontend representation as
$R=(R^{\mathrm{tok}},R^{\mathrm{feat}})$.  The component
$R^{\mathrm{tok}}$ contains one or more ordered token streams whose elements
index tokenizer vocabularies or codebooks, while $R^{\mathrm{feat}}$ contains
one or more ordered streams of feature frames.  Either component may be empty,
but not both, and each stream retains its own temporal order and native frame
rate.  Besides linguistic content needed by the backend, these representations
may retain prosody, speaker characteristics, and channel conditions.  The
input segment $x_I$ spans a finite source-time interval $I=[t_0,t_0+\Delta)$,
where $t_0$ is the start time and $\Delta$ is the segment duration. The adversary
intercepts only the token and feature streams produced when the frontend
processes this segment. We denote this fragmented frontend representation by
\begin{equation}
 R_I^{\mathrm{frag}}(x)=T(x_I)
 =\bigl(R_I^{\mathrm{tok}}(x),R_I^{\mathrm{feat}}(x)\bigr).
 \label{eq:fragmented-representation}
\end{equation}
Here, $R_I^{\mathrm{tok}}(x)$ and $R_I^{\mathrm{feat}}(x)$ are the token and
feature streams produced from $x_I$, respectively. The fragment retains the
visible streams and their temporal order but is not the full-utterance
representation $R$.

\subsubsection{Adversary knowledge and capabilities}
During one victim interaction, the adversary can read the fragmented
representation sent by the frontend, but cannot obtain its source waveform or
transcript, arbitrary user-device memory, or service-side backend states.  Read access may
arise from interception of readable representations at the client--service interface, a
compromised endpoint, an honest-but-curious service that legitimately receives
the same representation, or a third-party large-model gateway or relay that
terminates the user connection and forwards requests to a model provider.  The
adversary also has black-box access to the same frontend $T$: it may submit
chosen audio and observe the corresponding frontend representation, but cannot
read or modify the parameters of $T$.  It additionally possesses a public
auxiliary speech corpus with transcripts and may use its speaker metadata to
train an attack model offline on different utterances from the same auxiliary
speaker.  Victim speech, identity labels, and victim-specific training data
are unavailable.

Formally, let $v$ denote the victim and let $x_v^r$ denote the victim utterance
that produced the observed representation, where superscript $r$ marks the
representation source.  The adversary observes $R_I^{\mathrm{frag}}(x_v^r)$
during that interaction, but not $x_v^r$ itself or its transcript.  It has no
victim-specific training data or parameters of $G$ and does not use the
evaluation models for training or output selection. Its resources comprise
frontend queries, public auxiliary data, a finite leakage duration, and
attack-model capacity and training compute.

\subsubsection{Adversary objective}
An attack trial for victim $v$ additionally contains an enrollment utterance
$x_v^e$ disjoint from $x_v^r$ and available only to the experimenter, where
superscript $e$ denotes enrollment, and
attacker-chosen target text $y$ independent of the source content.  Given the
fragmented representation and target text, an attack algorithm $A$ generates
$\hat{x}=A(R_I^{\mathrm{frag}}(x_v^r),y)$. The attack aims to preserve the
victim's vocal identity, convey the chosen text, and produce natural speech.
We describe these goals through real-valued scores.

Let $S$ be an independent speaker-similarity scorer, $C$ a fixed ASR
transcription map, $d$ the normalized edit distance relative to the target
text, and $Q$ a quality scorer that predicts speech naturalness. For the same
generated waveform, we define identity similarity $s_{\mathrm{id}}$, content
error $e_{\mathrm{text}}$, and predicted naturalness $s_{\mathrm{qual}}$:
\begin{align}
 s_{\mathrm{id}}=S(\hat{x},x_v^e),
 e_{\mathrm{text}}=d(C(\hat{x}),y),
 s_{\mathrm{qual}}=Q(\hat{x}).
 \label{eq:trial-scores}
\end{align}

Subject to the stated knowledge, capability, and resource constraints, the
adversary seeks higher identity similarity and predicted naturalness together
with lower content error. The corresponding multi-objective formulation is
given below, with expectations over the prespecified trial distribution of
victims, source and enrollment utterances, interception intervals, target texts,
and attack randomness. The three components specify preferred directions and
allow tradeoffs rather than being combined into a weighted aggregate:
\begin{equation}
 \underset{A}{\operatorname{maximize}}\quad
 \bigl(\mathbb E[s_{\mathrm{id}}],\,-\mathbb E[e_{\mathrm{text}}],\,
       \mathbb E[s_{\mathrm{qual}}]\bigr).
 \label{eq:attack-objective}
\end{equation}

Equation~\eqref{eq:attack-objective} characterizes three system-level attack
goals rather than prescribing a direct training loss. The formulation
emphasizes that impersonation involves preserving vocal identity together with
conveying attacker-chosen content and retaining natural-speech quality, with
possible tradeoffs among these goals.

\subsection{Motivation}

Direct voice cloning from reference audio is well studied. Few- and zero-shot
systems can extract a speaker condition from a short reference recording and
synthesize new textual
content~\cite{arik2018cloning,jia2018transfer,casanova2022yourtts,wang2023valle}.
Consistently, speaker-verification research shows that learned speech
representations encode identity information sufficient for speaker
matching~\cite{snyder2018xvectors,desplanques2020ecapa,wang2023campp}. More recent
evidence extends this privacy concern to general-purpose speech representations
and speech language models: self-supervised representations support age
estimation~\cite{truong2022speakerage}; HearSay demonstrates private-attribute
inference from acoustic voiceprints by audio language
models~\cite{wang2026hearsay}; and SpInv reveals recoverable voiceprint
information in end-to-end speech-language-model frontend
outputs~\cite{lu2026spinv}. These studies indicate that speech representations
may retain acoustic cues associated with speaker timbre and identity in
addition to linguistic content.

Together, this evidence motivates an unresolved security question: without
access to a victim reference waveform, can an adversary turn a single
intercepted short frontend representation into reusable speaker conditioning
for attacker-chosen speech independent of the source content? Reference-audio
cloning establishes that speaker conditioning can support synthesis of new
content, while embedding recovery establishes that identity information
persists in frontend representations; neither determines whether leaked
representations retain the acoustic detail required for direct speech
generation. This motivates learning to reuse such information across
utterances and testing whether short leaks support impersonation of unseen
speakers.

\begin{figure*}[!t]
  \centering
  \includegraphics[width=0.98\textwidth]{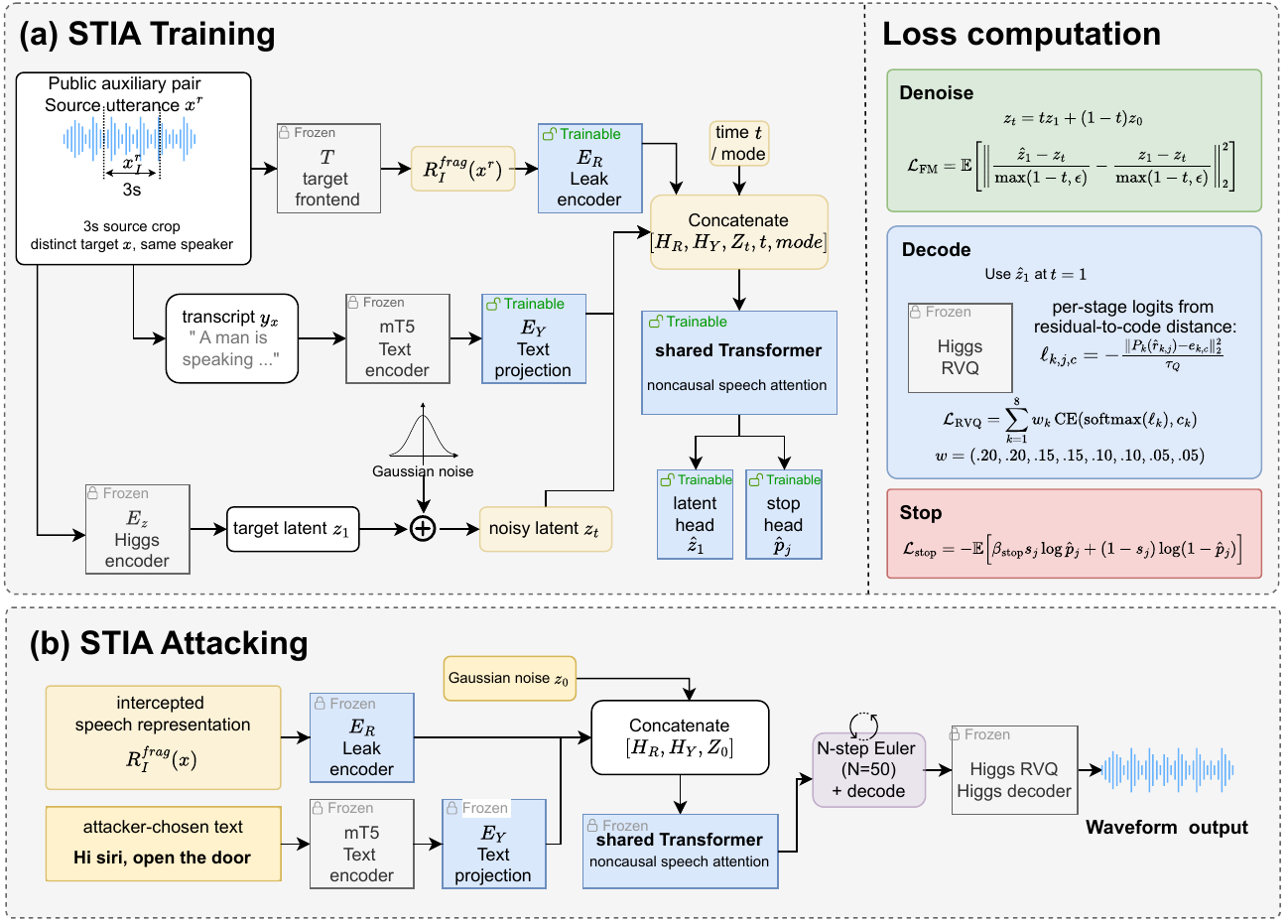}
  \caption{Overview of \stia. (a) Cross-utterance training with flow-matching,
  codec, and stopping losses. (b) Generation from leaked representations and
  chosen text via Euler sampling and codec decoding.}
  \label{fig:stia-overview}
\end{figure*}

\section{Speech Token Inversion Attack}
\label{sec:method}

\subsection{Overview}

\stia converts fragmented frontend representations into generation conditions
for attacker-chosen-text speech. As shown in Fig.~\ref{fig:stia-overview}, a
leak encoder maps the representation to a speaker-related condition sequence,
while a frozen mT5 encoder~\cite{xue2021mt5} and trainable text projection supply the target-content
condition. The two sequences are concatenated as inputs to a Transformer whose
parameters are shared between denoise and decode modes. It learns conditional
flow matching~\cite{lipman2023flow} in the continuous latent space of the Higgs
encoder~\cite{bosonai2026higgs}.
Its denoise mode learns to move from noise toward speech latents,
its decode mode provides codec-aware endpoint supervision, and a stop head
predicts the valid output boundary.

Figure~\ref{fig:stia-overview}(a) shows training on same-speaker, cross-utterance
pairs from public auxiliary speech: constructing leakage and text conditions
and supervising the generator with target speech latents. The target frontend
only supplies black-box query outputs and is excluded from gradient
computation; mT5 and the Higgs codec components remain frozen. The leak encoder,
text projection, shared Transformer, and its output heads are trained.
Figure~\ref{fig:stia-overview}(b) shows the fixed-model attack: encoding the
intercepted representation and new text, sampling a latent from Gaussian noise,
applying final decode, and synthesizing speech through frozen RVQ and the Higgs
waveform decoder. For readability, the figure omits latent-standardization bars
and abbreviates frame- and channel-averaged losses.

\subsection{STIA Training}
\label{sec:stia-training}

\noindent\textbf{Training pairs and conditioning.}
Each training example pairs target utterance $x$ with a distinct source
utterance $x^r$ from the same speaker, where superscript $r$ denotes the
representation source. A source crop $x_I^r$ over interval $I$ passes through
frozen frontend $T$ to produce $R_I^{\mathrm{frag}}(x^r)$; the target utterance
supplies transcript $y_x$ and acoustic supervision. Identity is thus shared
across the two utterances, whereas target text specifies the generated content.

The trainable leak encoder $E_R$ maps heterogeneous frontend outputs to a
condition sequence $H_R$ of a common width. Discrete tokens use codebook-specific
embeddings aggregated across codebooks; continuous features use a linear
projection and normalization. For mixed representations, the aligned branches
are concatenated along the feature dimension and projected to the shared width.
A validity mask preserves the temporal extent of the observed sequence:
\begin{equation}
 H_R=E_R(R_I^{\mathrm{frag}}(x^r)).
 \label{eq:training-conditions}
\end{equation}

The frozen multilingual mT5 encoder $M_{\mathrm{T5}}$ encodes the target
transcript, and trainable projection $E_Y$ maps it to a text-condition sequence
$H_Y$ of the same width. This permits a leaked condition to be combined with
different target texts, without requiring the target text to match the source
transcript:
\begin{equation}
 H_Y=E_Y(M_{\mathrm{T5}}(y_x)).
 \label{eq:text-condition}
\end{equation}

The shared Transformer assembles $H_R$, $H_Y$, speech latents, and time and mode
embeddings into one sequence. Condition positions cannot attend to speech
positions, whereas speech positions can attend to all valid condition and
speech positions. This noncausal speech attention combines identity cues,
target text, and full latent context. Cross-utterance supervision encourages
speaker information that can be reused in generating the target content,
but does not enforce complete disentanglement of identity and content.

\noindent\textbf{Latent flow matching.}
We use conditional flow matching~\cite{lipman2023flow}, adapting ELF's scheme
of denoising in a continuous embedding space followed by
decoding~\cite{hu2026elf} to speech-latent generation.
The frozen Higgs encoder $E_Z$ extracts the pre-quantization target latent
$z_1=E_Z(x)\in\mathbb{R}^{L\times d_Z}$, where $L$ and $d_Z$ are the valid
frame count and channel dimension. We standardize it using auxiliary-training-set
channelwise mean $\mu_Z$ and standard deviation $\sigma_Z$. The continuous
pre-quantization space admits interpolation paths between noise and the target:
\begin{equation}
 \bar z_1=\frac{z_1-\mu_Z}{\sigma_Z}.
 \label{eq:latent-normalization}
\end{equation}

Conditioned on $H_R$ and $H_Y$, the flow model learns the distribution of
standardized target latents. Denoise training samples noise $z_0$ from
$\mathcal{N}(0,4\mathbf I)$, where $\mathbf I$ is the identity matrix, and a
time $t$ from a time distribution $p_t$ on $(0,1)$. The state $z_t$ on the
linear path is
\begin{equation}
 z_t=(1-t)z_0+t\bar z_1.
 \label{eq:probability-path}
\end{equation}

As in ELF~\cite{hu2026elf}, we use an \emph{x-prediction, v-loss} strategy:
the network predicts the clean
endpoint rather than velocity directly. Let $F_\theta$ denote the shared
Transformer mapping with its latent and stop heads, where $\theta$ collects
trainable parameters, $u$ is the noisy latent input, $\tau$ is the network's
time input, and $b$ selects denoise or decode mode. The auxiliary
self-conditioning endpoint $q$ is zero when unused. Its outputs are clean-endpoint
estimate $\hat{\bar z}_1$ and framewise stop probabilities $\hat p$. Denoise
mode uses $u=z_t$ and $\tau=t$:
\begin{equation}
 (\hat{\bar z}_1,\hat p)=F_\theta(H_R,H_Y,u,\tau,b;q).
 \label{eq:shared-model}
\end{equation}

The predicted endpoint is converted to velocity through the remaining path
length. Let $d_t=\max(1-t,\epsilon)$, where $\epsilon$ is a numerical-stability
floor. The predicted velocity $\hat v_t$ is
\begin{equation}
 \hat v_t=\frac{\hat{\bar z}_1-z_t}{d_t}.
 \label{eq:flow-velocities}
\end{equation}
The target velocity parameterization $v_t^*$ is computed identically from the
true endpoint. When the floor is inactive, it equals the exact linear-path
velocity $\bar z_1-z_0$; near the endpoint, the floor limits error amplification:
\begin{equation}
 v_t^*=\frac{\bar z_1-z_t}{d_t}.
 \label{eq:target-velocity}
\end{equation}

The denoise objective minimizes squared velocity error. Let $j$ index valid
latent frames and $\mathbb E$ denote expectation over auxiliary pairs, noise,
time, and valid frames. Averaging over latent channels gives a clean-endpoint
error weighted by $d_t^{-2}$. Thus, endpoint prediction is linked to iterative
denoising along the path through velocity supervision:
\begin{equation}
 \mathcal L_{\mathrm{FM}}=
 \mathbb E\left[\frac{\|\hat v_{t,j}-v^*_{t,j}\|_2^2}{d_Z}\right].
 \label{eq:flow-loss}
\end{equation}

\noindent\textbf{Codec-aware decoding.}
Flow matching supervises denoising in continuous space, whereas the resulting
speech also depends on discrete code selection by the frozen RVQ. Following
ELF's shared denoising/decoding design~\cite{hu2026elf}, decode mode adds
target-code supervision to connect the predicted endpoint to the Higgs
quantization structure. It shares the Transformer and output heads with denoise
mode and is not a separate waveform decoder.

Decode training constructs input $z_a$ using independent Gaussian noise
$\xi\sim\mathcal N(0,\mathbf I)$ and corruption coefficient $a$ sampled from
a distribution $p_a$ on $(0,1)$.
The network receives $u=z_a$, but its time is fixed to $\tau=1$ and its mode to
$b=\mathrm{decode}$:
\begin{equation}
 z_a=a\bar z_1+(1-a)\xi.
 \label{eq:decode-input}
\end{equation}
The predicted endpoint is denormalized before RVQ supervision:
\begin{equation}
 \hat z_1=\sigma_Z\cdot\hat{\bar z}_1+\mu_Z.
 \label{eq:latent-denormalization}
\end{equation}

Let $k\in\{1,\ldots,8\}$ index RVQ stages, $e_{k,c}$ denote codeword $c$ at
stage $k$, and $P_k$ its frozen input projection. Target indices $c^*_{k,j}$ are
obtained by quantizing $z_1$. The predicted residual $\hat r_{k,j}$ is
initialized with $\hat z_{1,j}$ and updated by subtracting the output projections
of preceding ground-truth codewords. This teacher forcing supervises each stage
under the same target-code prefix. Classification logits $\ell_{k,j,c}$ use
projected residual-to-codeword distances, with temperature $\tau_Q$:
\begin{equation}
 \ell_{k,j,c}=-\frac{\|P_k(\hat r_{k,j})-e_{k,c}\|_2^2}{\tau_Q}.
 \label{eq:teacher-forced-rvq}
\end{equation}

Let $\ell_{k,j}$ collect the logits over all codewords at a frame,
$\operatorname{CE}$ denote cross-entropy between predicted probabilities and the target code index,
and $w_k$ be normalized stage weights.
The RVQ loss weights stages and averages over decode examples and their valid
frames. Codebooks and projections remain frozen, but the distance logits
propagate gradients to the predicted latent:
\begin{equation}
 \mathcal L_{\mathrm{RVQ}}=
 \mathbb E\left[\sum_{k=1}^{8}w_k\operatorname{CE}(\operatorname{softmax}(\ell_{k,j}),c^*_{k,j})\right].
 \label{eq:rvq-loss}
\end{equation}

\noindent\textbf{Stopping and joint training.}
The stop head predicts stopping probability $\hat p_j$ at each frame. Let
$s_j$ be a binary label equal to $1$ only at the final valid frame, and
$\beta_{\mathrm{stop}}$ a positive-class balancing weight. Both modes use
weighted binary cross-entropy to learn output boundaries alongside acoustic
generation:
\begin{equation}
 \mathcal L_{\mathrm{stop}}=-\mathbb E\big[\beta_{\mathrm{stop}}s_j\log\hat p_j+(1-s_j)\log(1-\hat p_j)\big].
 \label{eq:stop-loss}
\end{equation}

Each example selects denoise mode with probability $0.8$ and decode mode with
probability $0.2$. The former learns continuous denoising, the latter learns
target-code consistency, and stopping supervision applies to both. The expected
training objective is therefore given below, with the stopping loss also
averaged over mode sampling:
\begin{equation}
 \mathcal L=0.8\mathcal L_{\mathrm{FM}}+0.2\mathcal L_{\mathrm{RVQ}}
             +0.1\mathcal L_{\mathrm{stop}}.
 \label{eq:joint-loss}
\end{equation}

As in ELF~\cite{hu2026elf}, self-conditioning supplies a preliminary endpoint
prediction as $q$ without
gradient propagation, providing an existing clean-latent estimate for iterative
prediction. It is enabled for a minibatch with probability $0.5$. Each example
jointly drops leakage and text conditions with probability $0.1$, denoting empty
conditions by $\varnothing$, to learn unconditional predictions for
classifier-free guidance at attack time~\cite{ho2022cfg}.

Algorithm~\ref{alg:stia-training} summarizes five training stages: constructing
cross-utterance conditions and targets, sampling denoise or decode mode and
forming noisy inputs, predicting with the shared network, computing the
mode-specific primary loss and stopping loss, and aggregating losses to update
attack parameters. Here,
$\mathcal B$ denotes a minibatch, and $\ell_{\mathrm{FM},j}$,
$\ell_{\mathrm{RVQ},j}$, and $\ell_{\mathrm{stop},j}$ denote per-frame loss terms,
with the inactive mode's loss set to zero. $\mathcal L_{\mathcal B}$ averages
the sum of primary losses plus $0.1$ times the stopping loss over all valid
frames in the minibatch; mode probabilities are realized through sampling
rather than applied again as weights.

\begin{algorithm}[t]
\caption{Training the \stia attack model}
\label{alg:stia-training}
\footnotesize
\begin{algorithmic}[1]
\Statex \textbf{Input:} Same-speaker triples $(x,x^r,y_x)$ with $x\ne x^r$;
\Statex \hspace{\algorithmicindent} black-box $T$; frozen $M_{\mathrm{T5}},E_Z$, RVQ; $p_t,p_a$
\Statex \textbf{Output:} Trained $\theta$ and statistics $(\mu_Z,\sigma_Z)$
\State Initialize $\theta$; estimate $\mu_Z,\sigma_Z$ from $E_Z(x)$
\For{each minibatch $\mathcal B$}
  \State Initialize an empty frame-loss list $\mathcal S$
  \State Enable self-conditioning with probability $0.5$
  \ForAll{$(x,x^r,y_x)\in\mathcal B$}
    \Statex \hspace{\algorithmicindent}\hspace{\algorithmicindent}$\blacktriangleright$ \textbf{1. Conditions and targets}
    \State Crop $x_I^r$; encode leakage as $H_R\gets E_R(T(x_I^r))$
    \State Encode target text as $H_Y\gets E_Y(M_{\mathrm{T5}}(y_x))$
    \State $z_1\gets E_Z(x)$; $\bar z_1\gets(z_1-\mu_Z)/\sigma_Z$
    \State Set $s_L=1$; all other stop labels $s_j=0$
    \State Jointly drop $H_R,H_Y$ with probability $0.1$
    \Statex \hspace{\algorithmicindent}\hspace{\algorithmicindent}$\blacktriangleright$ \textbf{2. Add noise: mode-specific inputs}
    \State Sample $b\in\{\mathrm{denoise},\mathrm{decode}\}$ with $0.8/0.2$
    \If{$b=\mathrm{denoise}$}
      \State Sample $z_0,t$; form $u\gets z_t$, $\tau\gets t$ via Eq.~\eqref{eq:probability-path}
    \Else
      \State Sample $\xi,a$; form $u\gets z_a$, $\tau\gets1$ via Eq.~\eqref{eq:decode-input}
    \EndIf
    \Statex \hspace{\algorithmicindent}\hspace{\algorithmicindent}$\blacktriangleright$ \textbf{3. Denoise / decode: shared prediction}
    \State $q\gets0$
    \If{self-conditioning is enabled}
      \State $(q,\_)\gets F_\theta(H_R,H_Y,u,\tau,b;0)$; detach $q$
    \EndIf
    \State $(\hat{\bar z}_1,\hat p)\gets F_\theta(H_R,H_Y,u,\tau,b;q)$
    \Statex \hspace{\algorithmicindent}\hspace{\algorithmicindent}$\blacktriangleright$ \textbf{4. Losses: flow, codec, and stopping}
    \State $\ell_{\mathrm{FM},j},\ell_{\mathrm{RVQ},j}\gets0,0$ for valid frames $j$
    \If{$b=\mathrm{denoise}$}
      \State Compute per-frame $\ell_{\mathrm{FM},j}$ via Eq.~\eqref{eq:flow-loss}
    \Else
      \State Denormalize $\hat{\bar z}_1$; quantize $z_1$ to target codes
      \State Compute per-frame $\ell_{\mathrm{RVQ},j}$ via Eqs.~\eqref{eq:teacher-forced-rvq}--\eqref{eq:rvq-loss}
    \EndIf
    \State Compute weighted stop BCE $\ell_{\mathrm{stop},j}$ via Eq.~\eqref{eq:stop-loss}
    \State Append per-frame $\ell_{\mathrm{FM},j}+\ell_{\mathrm{RVQ},j}+0.1\ell_{\mathrm{stop},j}$ to $\mathcal S$
  \EndFor
  \Statex \hspace{\algorithmicindent}$\blacktriangleright$ \textbf{5. Joint loss and parameter update}
  \State Average frame losses: $\mathcal L_{\mathcal B}\gets\operatorname{Mean}(\mathcal S)$
  \State Update only $\theta$ with AdamW using $\nabla_\theta\mathcal L_{\mathcal B}$
\EndFor
\end{algorithmic}
\end{algorithm}

\subsection{STIA Attacking}
\label{sec:stia-attacking}

At attack time, fixed encoders transform $R_I^{\mathrm{frag}}(x_v^r)$
intercepted from victim $v$ and chosen text $y$ into conditions $H_R,H_Y$.
No victim waveform, source transcript, or enrollment speech is required.
Let $L_{\max}$ be the latent-length limit determined from the chosen text.
Initial state $z^{(0)}\in\mathbb R^{L_{\max}\times d_Z}$ is sampled from the
same Gaussian base distribution as denoise training:
\begin{equation}
 z^{(0)}\sim\mathcal N(0,4\mathbf I).
 \label{eq:attack-initialization}
\end{equation}

At sampling step $n$, the model predicts clean endpoints with $(H_R,H_Y)$ and
with empty conditions, denoted $\hat{\bar z}_{1,n}^{\mathrm c}$ and
$\hat{\bar z}_{1,n}^{\mathrm u}$; superscripts $\mathrm c$ and $\mathrm u$
mean conditional and unconditional. Classifier-free guidance
(CFG)~\cite{ho2022cfg} combines the predictions with scale $s$ to strengthen
adherence to leakage and text conditions. Each path uses its own preceding endpoint as
self-conditioning, initialized to zero:
\begin{equation}
 \hat{\bar z}_{1,n}^{\mathrm{cfg}}=\hat{\bar z}_{1,n}^{\mathrm u}
 +s\big(\hat{\bar z}_{1,n}^{\mathrm c}-\hat{\bar z}_{1,n}^{\mathrm u}\big).
 \label{eq:cfg-endpoint}
\end{equation}

The guided endpoint defines a velocity from the current noisy state toward
conditional speech. Let $t_n$ denote the current time, $\epsilon_{\mathrm{sam}}$
the sampling denominator floor, and $v_n$ the velocity evaluated at the
current state with its self-conditioning:
\begin{equation}
 v_n=\frac{\hat{\bar z}_{1,n}^{\mathrm{cfg}}-z^{(n)}}
               {\max(1-t_n,\epsilon_{\mathrm{sam}})}.
 \label{eq:inference-field}
\end{equation}

We integrate the learned flow with an explicit Euler ordinary differential
equation (ODE) sampler, as in ELF's ODE sampling scheme~\cite{hu2026elf}.
We use $N=50$ steps in all experiments. Integration ends at $t_\star=0.99$ on the uniform grid
$t_n=nt_\star/N$, where $n=0,\ldots,N$, with step size $\Delta t=t_\star/N$.
Each update advances the latent along the predicted velocity:
\begin{equation}
 z^{(n+1)}=z^{(n)}+\Delta t\,v_n.
 \label{eq:euler-update}
\end{equation}

After sampling, the model performs one conditional decode at time $1$ on
$z^{(N)}$, without CFG and retaining the last conditional endpoint as
self-conditioning. This step produces the final latent and stop probabilities.
Valid length $L$ is the first frame after $L_{\max}/2$ whose stop probability
reaches threshold $\tau_{\mathrm{stop}}$, or $L_{\max}$ if none does. The
prediction is then truncated to $L$ frames and denormalized.

Frozen RVQ converts the final continuous latent into discrete codes. Let
$r_{k,j}$ denote the inference residual, initialized with
$r_{1,j}=\hat z_{1,j}$, $\hat c_{k,j}$ the selected code index, and $O_k$ the
frozen output projection at stage $k$. Each stage
selects the nearest codeword to its projected residual, then updates the
residual by subtracting that codeword's output projection. Unlike training
teacher forcing, these updates use only predicted codewords:
\begin{equation}
 \hat c_{k,j}=\arg\min_c\|P_k(r_{k,j})-e_{k,c}\|_2^2.
 \label{eq:inference-rvq}
\end{equation}

Let $\hat C\in\mathbb N^{8\times L}$ be the array of predicted code indices
$\hat c_{k,j}$ ordered by stage and frame, $Q_Z$ denote frozen RVQ reconstruction from indices to
quantized latents, and $D_Z$ be the Higgs waveform decoder. Decoding the
reconstructed latent yields the attack waveform:
\begin{equation}
 \hat x=D_Z(Q_Z(\hat C)).
 \label{eq:attack-waveform}
\end{equation}

Algorithm~\ref{alg:stia-attack} summarizes the complete attack in four stages.
First, it encodes the leaked representation and target text and initializes
latents with Gaussian noise. Second, it iteratively performs Euler ODE sampling
using CFG to combine conditional and unconditional predictions. Third, a final
decode determines the valid output length and denormalizes the latent.
Fourth, frozen RVQ and the waveform decoder synthesize speech. These stages
convert a single short representation leak into speech conveying
attacker-chosen text.

\begin{algorithm}[t]
\caption{Attacking with \stia: from leaked conditions to speech}
\label{alg:stia-attack}
\footnotesize
\begin{algorithmic}[1]
\Statex \textbf{Input:} Intercepted $R_I^{\mathrm{frag}}(x_v^r)$, chosen text $y$;
\Statex \hspace{\algorithmicindent} fixed models, $(\mu_Z,\sigma_Z)$; $N=50$, $s=2$,
\Statex \hspace{\algorithmicindent} $\epsilon_{\mathrm{sam}}=0.01$, $\tau_{\mathrm{stop}}=0.5$
\Statex \textbf{Output:} Attack waveform $\hat x$
\Statex $\blacktriangleright$ \textbf{1. Conditions and noise initialization}
\State Encode leakage as $H_R\gets E_R(R_I^{\mathrm{frag}}(x_v^r))$
\State Encode text as $H_Y\gets E_Y(M_{\mathrm{T5}}(y))$; estimate $L_{\max}$
\State Sample $L_{\max}$ noise frames $z^{(0)}$ using Eq.~\eqref{eq:attack-initialization}
\State Set $t_n\gets0.99n/N$ and $\Delta t\gets0.99/N$
\State Initialize separate self-conditions $q^{\mathrm c},q^{\mathrm u}\gets0,0$
\Statex $\blacktriangleright$ \textbf{2. Denoise: guided Euler sampling}
\For{$n=0,\ldots,N-1$}
  \State Set $(u,\tau,b)\gets(z^{(n)},t_n,\mathrm{denoise})$
  \State Predict $\hat{\bar z}_{1,n}^{\mathrm c}$ with $(H_R,H_Y,q^{\mathrm c})$ via Eq.~\eqref{eq:shared-model}
  \State Predict $\hat{\bar z}_{1,n}^{\mathrm u}$ with $(\varnothing,\varnothing,q^{\mathrm u})$ via Eq.~\eqref{eq:shared-model}
  \State Combine endpoints with CFG using Eq.~\eqref{eq:cfg-endpoint}
  \State Compute velocity $v_n$ using Eq.~\eqref{eq:inference-field}
  \State Update $z^{(n+1)}$ with the Euler step in Eq.~\eqref{eq:euler-update}
  \State $q^{\mathrm c}\gets\hat{\bar z}_{1,n}^{\mathrm c}$; $q^{\mathrm u}\gets\hat{\bar z}_{1,n}^{\mathrm u}$
\EndFor
\Statex $\blacktriangleright$ \textbf{3. Decode: final latent and stopping}
\State Set $(u,\tau,b)\gets(z^{(N)},1,\mathrm{decode})$
\State Predict $(\hat{\bar z}_1,\hat p)$ via Eq.~\eqref{eq:shared-model} with $(H_R,H_Y,q^{\mathrm c})$
\Statex \hspace{\algorithmicindent}using only the conditional prediction, without CFG
\State $L\gets\min\{j:j>L_{\max}/2,\ \hat p_j\geq\tau_{\mathrm{stop}}\}$
\Statex \hspace{\algorithmicindent} Use $L_{\max}$ if no such frame exists
\State $\hat z_1\gets(\sigma_Z\cdot\hat{\bar z}_1+\mu_Z)_{1:L}$
\Statex $\blacktriangleright$ \textbf{4. RVQ and waveform synthesis}
\State Quantize $\hat z_1$ to $\hat C$ with 8-stage RVQ, Eq.~\eqref{eq:inference-rvq}
\Statex \hspace{\algorithmicindent}updating residuals using predicted codes only
\State Synthesize $\hat x$ with the frozen Higgs decoder, Eq.~\eqref{eq:attack-waveform}
\State \Return $\hat x$
\end{algorithmic}
\end{algorithm}

\section{Experiments}
\label{sec:evaluation}

\subsection{Experimental Setup}
\label{sec:setup}

\noindent\textbf{Datasets.}
Our training and validation sets are two disjoint subsets of the public Emilia
dataset partitioned by speaker ID, with no shared
speakers~\cite{he2024emilia}. Each experiment uses data from one specific
language within Emilia for both training and validation. Main experiments use
Emilia-English, while separate Chinese experiments use Emilia-Chinese. We
train an attack model separately for each language and evaluate it on the
validation subset of the same language; we do not perform cross-language
validation.

\noindent\textbf{Target models.}
We select Moshi, Kimi-Audio, and Qwen3-Omni as three representative target
models and use only their input-side audio
frontends~\cite{defossez2024moshi,kimiteam2025kimiaudio,xu2025qwen3omni}.
Access to all three targets follows a black-box input--output interface:
submitting audio and receiving frontend representations, without access to
model parameters or gradients. Representations of auxiliary speech are
obtained through frontend queries. The target models remain fixed and are
excluded from attack-model parameter optimization and backpropagation;
their backends are not involved in attack training.

Table~\ref{tab:frontends} summarizes the audio frontend architectures,
processing pipelines, and output formats of the three target models. The
selected representations are taken after Mimi RVQ, at Kimi-Audio's two branch
outputs, and at the AuT output, respectively, to instantiate the assumed
split-deployment interfaces. The attacker receives only victim representations
at these interfaces, not raw waveforms. Qwen3-Omni's Talker output codec is not
part of the selected input-side audio frontend.

\begin{table*}[t]
\caption{Audio frontend architectures, processing pipelines, output formats, and representation dimensions of Moshi, Kimi-Audio, and Qwen3-Omni at the input-side interfaces evaluated in the split-deployment attack setting.}
\label{tab:frontends}
\centering
\footnotesize
\setlength{\tabcolsep}{4pt}
\renewcommand{\arraystretch}{1.18}
\begin{tabular}{
  >{\raggedright\arraybackslash}p{0.72in}
  >{\raggedright\arraybackslash}p{1.14in}
  >{\raggedright\arraybackslash}p{2.52in}
  >{\raggedright\arraybackslash}p{1.18in}
  >{\raggedright\arraybackslash}p{0.78in}}
\toprule
\textbf{Target model} & \textbf{Frontend model} & \textbf{Frontend processing pipeline} & \textbf{Frontend output format} & \textbf{Output dimension} \\
\midrule
\textbf{Moshi} &
Mimi neural audio codec &
24-kHz mono waveform $\rightarrow$ causal SEANet encoder (strides $4,5,6,8,2$) $\rightarrow$ 8-layer, 8-head Transformer $\rightarrow$ 8-stage split RVQ &
Frame-synchronous discrete RVQ code-index streams at 12.50 Hz &
8 scalar IDs/frame; 2,048 entries per stage \\
\addlinespace[2pt]
\textbf{Kimi-Audio} &
GLM-4-Voice semantic tokenizer and adapted Whisper encoder &
16-kHz mono waveform $\rightarrow$ 128-bin log-Mel features $\rightarrow$ parallel semantic and acoustic branches: 32-layer tokenizer with single-codebook VQ; four 1,280-D Whisper frames with a 2-layer adaptor &
Aligned semantic ID and continuous acoustic feature at 12.50 Hz &
1 scalar ID $+$ 3,584-D vector/frame \\
\addlinespace[2pt]
\textbf{Qwen3-Omni} &
Audio Transformer (AuT) encoder &
16-kHz mono waveform $\rightarrow$ 128-bin log-Mel features (25-ms window, 10-ms hop) $\rightarrow$ three stride-2 Conv2D blocks $\rightarrow$ 32-layer, 650M-parameter AuT (width 1,280) &
Frame-synchronous continuous AuT hidden sequence at 12.50 Hz &
2,048-D vector/frame \\
\bottomrule
\end{tabular}
\end{table*}

\noindent\textbf{Comparison baseline.}
We combine SpInv~\cite{lu2026spinv}, the closest representation-inversion method,
with the SpeechT5 synthesizer~\cite{ao2022speecht5} to construct a SpInv+TTS
baseline for this task. SpInv supports attacker-specified embedding spaces;
we retrain it specifically in the 512-dimensional x-vector space used by
SpeechT5 so that the recovered vectors are compatible with the synthesizer.
SpeechT5 then generates speech conditioned on the predicted embedding and
attacker-chosen text; neither stage receives genuine victim audio. Both attacks
use the same leakage segments, target texts, victim trials, and evaluators.
The comparison evaluates two-stage embedding recovery and synthesis against
\stia's direct representation-conditioned generation. Evaluation models
participate in neither training nor output selection.

\noindent\textbf{Metrics.}
\emph{CAM++ similarity} (CAM in tables) is computed by extracting speaker
embeddings from generated speech and the victim's separate genuine enrollment
speech using CAM++, then taking their cosine similarity: the embedding dot
product divided by the product of their Euclidean norms~\cite{wang2023campp}.
It measures preservation of the victim's vocal identity; higher is better.

\emph{ECAPA-TDNN similarity} (ECAPA in tables) is computed from ECAPA-TDNN
embeddings of generated speech and the same genuine enrollment speech, taking
their cosine similarity as the dot product divided by the product of their
norms~\cite{desplanques2020ecapa}. It measures identity preservation with a
second speaker encoder; higher is better.

\emph{Word error rate} (WER) aligns the ASR transcript of generated English
speech with the attacker-chosen text, divides the minimum total number of word
substitutions, deletions, and insertions by the target word count, and reports
the result as a percentage. It measures how accurately the generated English
content matches the chosen text; lower is better.

\emph{Character error rate} (CER) aligns the ASR transcript of generated Chinese
speech with the attacker-chosen text, divides the minimum total number of
character substitutions, deletions, and insertions by the target character
count, and reports the result as a percentage. It measures how accurately the
generated Chinese content matches the chosen text; lower is better.

\emph{UTMOS} passes each generated waveform through a pretrained scoring model
to predict its mean opinion score without reference
audio~\cite{saeki2022utmos}. It estimates the perceptual naturalness of generated
speech; higher is better.

\noindent\textbf{Implementation details.}\label{sec:implementation}
M, L, and XL use 8, 12, and 16 layers, respectively; the default is XL (0.60B).
The main Moshi setting uses 3-second leaks and the largest Emilia-English
budget. Tables without a target-model column use Moshi and reuse its
main-configuration results. Training settings and cost appear in
Sec.~\ref{sec:hyperparameters-runtime}.

Leakage and enrollment utterances are disjoint; frozen target prompts are
independent of leaked content. Each configuration generates four samples per
recording for 1,000 recordings using 50-step Euler sampling and CFG scale 2.
All 4,000 utterance-level scores are aggregated separately by language.
Table~\ref{tab:main-results} reports mean $\pm$ standard deviation;
Tables~\ref{tab:language-ablation}--\ref{tab:conditioning-ablation} report means.
Reported statistics are rounded to two decimals.

\noindent\textbf{Result presentation.}
Subscripts indicate the direction and magnitude of absolute mean changes,
computed before rounding. The reference is SpInv+TTS on the same frontend in
Table~\ref{tab:main-results} and the first row of each table in
Tables~\ref{tab:language-ablation}--\ref{tab:conditioning-ablation}.
Error-rate differences are expressed in percentage points.
Bold marks the numerically best reported mean, determined before rounding,
within each frontend in Table~\ref{tab:main-results} and within each column in
Tables~\ref{tab:language-ablation}--\ref{tab:conditioning-ablation}.
Underlined changes in Table~\ref{tab:main-results} identify the largest
improvement across frontends for each metric.

Evaluators use 3D-Speaker's CAM++ and ECAPA-TDNN
checkpoints~\cite{chen2025threedspeaker}, OpenAI's Whisper-large-v3-turbo, and
UTMOS22-strong~\cite{saeki2022utmos}. All methods share fixed checkpoints and
identical ASR decoding and text-normalization settings; evaluators participate
in neither training nor output selection.

\subsection{Comparison with SpInv+TTS}
\label{sec:main-results}

\begin{table*}[t]
\caption{Comparison of \stia and SpInv+TTS on unseen Emilia speakers across three speech frontends, using identical 3-second representation leaks and attacker-chosen English prompts.}
\label{tab:main-results}
\centering
\footnotesize
\setlength{\tabcolsep}{4pt}
\renewcommand{\arraystretch}{1.15}
\newcommand{\resultdelta}[4]{\ensuremath{{\mathbf{#1} \pm #2}_{(#3\,#4)}}}
\begin{tabular*}{\textwidth}{@{\extracolsep{\fill}}llllll@{}}
\toprule
\textbf{Target model} & \textbf{Attack} & \textbf{CAM} $\uparrow$ & \textbf{ECAPA} $\uparrow$ & \textbf{WER (\%)} $\downarrow$ & \textbf{UTMOS} $\uparrow$ \\
\midrule
\multirow{2}{*}{\textbf{Moshi}} & SpInv+TTS & \result{0.22}{0.11} & \result{0.26}{0.10} & \result{13.37}{44.99} & \result{2.91}{0.65} \\
 & \textbf{\stia} (ours) & \resultdelta{0.78}{0.03}{\uparrow}{0.56} & \resultdelta{0.79}{0.05}{\uparrow}{0.53} & \resultdelta{3.80}{12.60}{\downarrow}{\underline{9.57}} & \resultdelta{2.92}{0.17}{\uparrow}{0.01} \\
\midrule
\multirow{2}{*}{\textbf{Kimi-Audio}} & SpInv+TTS & \result{0.13}{0.12} & \result{0.20}{0.12} & \result{12.40}{32.40} & \result{2.87}{0.52} \\
 & \textbf{\stia} (ours) & \resultdelta{0.82}{0.03}{\uparrow}{\underline{0.69}} & \resultdelta{0.85}{0.05}{\uparrow}{\underline{0.65}} & \resultdelta{3.84}{11.25}{\downarrow}{8.55} & \resultdelta{2.97}{0.22}{\uparrow}{\underline{0.10}} \\
\midrule
\multirow{2}{*}{\textbf{Qwen3-Omni}} & SpInv+TTS & \result{0.35}{0.09} & \result{0.33}{0.10} & \result{10.56}{26.27} & \result{2.91}{0.37} \\
 & \textbf{\stia} (ours) & \resultdelta{0.75}{0.04}{\uparrow}{0.40} & \resultdelta{0.78}{0.06}{\uparrow}{0.45} & \resultdelta{4.61}{13.70}{\downarrow}{5.95} & \resultdelta{2.91}{0.21}{\uparrow}{<0.01} \\
\bottomrule
\end{tabular*}
\end{table*}

For each frontend, \stia and SpInv+TTS receive the same 3-second victim
representation fragment and attacker-chosen English text and are evaluated
against the same genuine enrollment utterance. Table~\ref{tab:main-results}
shows higher similarity under both speaker encoders and lower mean WER for
\stia on all three frontends. Kimi-Audio yields the highest identity similarity
with mean WER of 3.84\%, close to Moshi's 3.80\%. UTMOS means remain
similar between the two attacks; the principal observed gains are in identity
similarity and text accuracy.

\subsection{Language Experiments}
\label{sec:language-results}

We conduct separate experiments on Emilia-English and Emilia-Chinese,
training architecturally identical \stia models and evaluating each language
with its own speakers and frozen prompt set. The experiments do not share
victim speakers but use the same Moshi frontend, model size, training steps,
and optimization settings. Table~\ref{tab:language-ablation} shows that the
attack retains speaker similarity in both languages. The Chinese group has
higher identity scores but lower UTMOS; with different speaker populations,
these differences cannot be attributed to language alone. English WER and
Chinese CER also use different counting units.

\begin{table}[t]
\caption{Separate English and Chinese evaluations of \stia with language-specific training on Moshi.}
\label{tab:language-ablation}
\centering
\footnotesize
\setlength{\tabcolsep}{2pt}
\renewcommand{\arraystretch}{1.15}
\begin{tabular*}{\columnwidth}{@{\extracolsep{\fill}}lcccc@{}}
\toprule
Language & CAM $\uparrow$ & ECAPA $\uparrow$ & \shortstack{WER/CER(\%) $\downarrow$} & UTMOS $\uparrow$ \\
\midrule
\textbf{English} & 0.78 & 0.79 & \textbf{3.80} & \textbf{2.92} \\
\addlinespace[3pt]
Chinese & \meandelta{\mathbf{0.82}}{\uparrow}{0.04} & \meandelta{\mathbf{0.84}}{\uparrow}{0.05} & \meandelta{4.05}{\uparrow}{0.25} & \meandelta{2.62}{\downarrow}{0.30} \\
\bottomrule
\end{tabular*}
\end{table}

\subsection{Auxiliary-Data Ablation}
\label{sec:data-ablation}

The auxiliary-data ablation compares three nested Emilia-English budgets,
selected with a fixed random seed and retaining valid
same-speaker, cross-utterance pairs.
Filtering, frontend, English prompts, leakage duration, and model size remain
unchanged, and all configurations are trained for 150,000 steps.
In Table~\ref{tab:data-ablation}, M denotes million clips and k denotes thousand
hours. The table shows that additional auxiliary data improves both
identity scores and lowers mean WER at a fixed update count, while UTMOS remains
largely unchanged. Identity gains from the middle to the largest budget are
smaller than those from the smallest to the middle budget.

\begin{table}[t]
\caption{Effect of auxiliary Emilia-English data budgets on \stia speech generation using the Moshi frontend.}
\label{tab:data-ablation}
\centering
\footnotesize
\setlength{\tabcolsep}{1pt}
\renewcommand{\arraystretch}{1.15}
\begin{tabular*}{\columnwidth}{@{\extracolsep{\fill}}cccccc@{}}
\toprule
Clips(M) & Dura.(h) & CAM $\uparrow$ & ECAPA $\uparrow$ & WER (\%) $\downarrow$ & UTMOS $\uparrow$ \\
\midrule
0.80 & 2.9k & 0.70 & 0.76 & 8.98 & \textbf{2.93} \\
\addlinespace[3pt]
1.60 & 5.9k & \meandelta{0.76}{\uparrow}{0.06} & \meandelta{0.78}{\uparrow}{0.03} & \meandelta{5.19}{\downarrow}{3.78} & \meandelta{2.92}{\downarrow}{<0.01} \\
\addlinespace[3pt]
\textbf{2.40} & 8.9k & \meandelta{\mathbf{0.78}}{\uparrow}{0.08} & \meandelta{\mathbf{0.79}}{\uparrow}{0.03} & \meandelta{\mathbf{3.80}}{\downarrow}{5.17} & \meandelta{2.92}{\downarrow}{<0.01} \\
\bottomrule
\end{tabular*}
\end{table}

\subsection{Leakage-Duration Ablation}
\label{sec:leakage-ablation}

We extract frontend representations from nested 2-, 3-, 5-, and 8-second
speech crops of the same victim source utterance. Only utterances long enough
for the largest crop are eligible; victims, English prompts, enrollment
utterances, and attack checkpoints are fixed across durations.
Table~\ref{tab:leakage-ablation} shows substantial identity similarity even
with the shortest leak, with both identity means increasing with duration.
Mean WER decreases from 2 to 3 seconds and then remains approximately stable;
UTMOS has no monotonic trend. The 3-second configuration therefore captures
most of the text-accuracy benefit in this comparison, while longer leaks
primarily yield additional identity similarity.

\begin{table}[t]
\caption{Effect of leaked representation duration on \stia speech generation using the Moshi frontend.}
\label{tab:leakage-ablation}
\centering
\footnotesize
\setlength{\tabcolsep}{2pt}
\renewcommand{\arraystretch}{1.15}
\begin{tabular*}{\columnwidth}{@{\extracolsep{\fill}}ccccc@{}}
\toprule
Leakage & CAM $\uparrow$ & ECAPA $\uparrow$ & WER (\%) $\downarrow$ & UTMOS $\uparrow$ \\
\midrule
2.00 s & 0.76 & 0.79 & 6.24 & 2.91 \\
\addlinespace[3pt]
\textbf{3.00 s} & \meandelta{0.78}{\uparrow}{0.03} & \meandelta{0.79}{\uparrow}{<0.01} & \meandelta{\mathbf{3.80}}{\downarrow}{2.44} & \meandelta{\mathbf{2.92}}{\uparrow}{0.01} \\
\addlinespace[3pt]
5.00 s & \meandelta{0.79}{\uparrow}{0.03} & \meandelta{0.81}{\uparrow}{0.03} & \meandelta{3.85}{\downarrow}{2.38} & \meandelta{2.90}{\downarrow}{<0.01} \\
\addlinespace[3pt]
8.00 s & \meandelta{\mathbf{0.80}}{\uparrow}{0.04} & \meandelta{\mathbf{0.82}}{\uparrow}{0.03} & \meandelta{3.95}{\downarrow}{2.29} & \meandelta{2.90}{\downarrow}{<0.01} \\
\bottomrule
\end{tabular*}
\end{table}

\subsection{Model-Scale Ablation}
\label{sec:model-scale}

The model-scale ablation compares M, L, and XL configurations with
approximately 0.35B, 0.50B, and 0.60B parameters, respectively.
Training data, leakage fragments, the Moshi frontend, English trial lists,
and other training settings are fixed.
Table~\ref{tab:model-scale} shows improvements in measured identity similarity,
text accuracy, and predicted naturalness as model size increases. XL has the
best mean on all four metrics within the tested range and is used as the
main configuration.

\begin{table}[t]
\caption{Effect of \stia model size on speech generation from 3-second Moshi representation leaks.}
\label{tab:model-scale}
\centering
\footnotesize
\setlength{\tabcolsep}{2pt}
\renewcommand{\arraystretch}{1.15}
\begin{tabular*}{\columnwidth}{@{\extracolsep{\fill}}lccccc@{}}
\toprule
Size & Params & CAM $\uparrow$ & ECAPA $\uparrow$ & WER (\%) $\downarrow$ & UTMOS $\uparrow$ \\
\midrule
\textbf{XL} & 0.60B & \textbf{0.78} & \textbf{0.79} & \textbf{3.80} & \textbf{2.92} \\
\addlinespace[3pt]
L & 0.50B & \meandelta{0.75}{\downarrow}{0.03} & \meandelta{0.77}{\downarrow}{0.02} & \meandelta{4.71}{\uparrow}{0.91} & \meandelta{2.82}{\downarrow}{0.10} \\
\addlinespace[3pt]
M & 0.35B & \meandelta{0.70}{\downarrow}{0.08} & \meandelta{0.72}{\downarrow}{0.07} & \meandelta{7.72}{\uparrow}{3.92} & \meandelta{2.59}{\downarrow}{0.33} \\
\bottomrule
\end{tabular*}
\end{table}

\subsection{Speaker Anonymization}
\label{sec:defense}

We use the VoicePrivacy 2022 B2 McAdams anonymization method and default
parameter settings as an input-side defense~\cite{patino2021mcanonymization}.
Each 3-second source crop is converted to 16-kHz mono audio and processed
using 20-ms windows, a 10-ms hop, and order-20 linear predictive coding (LPC).
The coefficient $\alpha\sim\mathcal U(0.5,0.9)$ is drawn once per speaker and
held fixed across that speaker's crops and repeated generation trials.
Boundary zero-padding covers the full crop; frontend representations are
then re-extracted from the anonymized waveform to condition \stia.

The attacker is trained on unanonymized auxiliary speech without retraining
or tuning; its checkpoint, sampling settings, and output selection rule remain
fixed. The anonymization transform and coefficients are unknown to the
attacker. Both input conditions use the same Moshi frontend, victims, and
English prompts; identity evaluation uses disjoint, genuine enrollment speech
of the victim's original identity.

Table~\ref{tab:defense} shows lower mean similarity under both speaker encoders
and lower UTMOS after anonymization, with a small increase in mean WER.
These results support attenuation
of this attacker, but distribution shift and reduced naturalness preclude
interpreting them as evidence that identity information has been removed;
adaptive attackers remain unevaluated.

\begin{table}[t]
\caption{Effect of McAdams input anonymization on a fixed \stia attacker using the Moshi frontend.}
\label{tab:defense}
\centering
\footnotesize
\setlength{\tabcolsep}{2pt}
\renewcommand{\arraystretch}{1.15}
\begin{tabular*}{\columnwidth}{@{\extracolsep{\fill}}lcccc@{}}
\toprule
Input & CAM $\uparrow$ & ECAPA $\uparrow$ & WER (\%) $\downarrow$ & UTMOS $\uparrow$ \\
\midrule
\textbf{Original} & \textbf{0.78} & \textbf{0.79} & \textbf{3.80} & \textbf{2.92} \\
\addlinespace[3pt]
Anonymized & \meandelta{0.44}{\downarrow}{0.35} & \meandelta{0.43}{\downarrow}{0.36} & \meandelta{4.51}{\uparrow}{0.71} & \meandelta{2.43}{\downarrow}{0.49} \\
\bottomrule
\end{tabular*}
\end{table}

\subsection{Conditioning and Text Controls}
\label{sec:conditioning-ablation}

We examine leakage and text conditioning through three input interventions:
No-leak removes leakage, No-text removes text, and No-leak-and-text removes
both. These interventions use the fixed main-configuration model without
retraining. Let $v$ denote the victim speaker and define
$R_v=R_I^{\mathrm{frag}}(x_v^r)$. Generation seeds and other settings are fixed
for these three interventions; Standard in
Table~\ref{tab:conditioning-ablation} reuses the main-experiment results.

No-leak removes all leakage-conditioning streams, retaining text in the
conditional branch while preserving normal behavior of the CFG unconditional
branch; the final decode also receives no leakage.  Training jointly drops
text and leakage, so the text-only no-leak condition may be out of distribution;
degradation alone cannot establish the necessity of leakage conditioning.

No-text retains leakage conditioning but supplies no target text, whereas
No-leak-and-text (Neither in Table~\ref{tab:conditioning-ablation}) supplies
neither leakage conditioning nor target text.
WER is not reported for these two conditions because no target text is supplied.

Mismatched text (Wrong text in Table~\ref{tab:conditioning-ablation}) is a
separate scoring negative control: the fixed model
generates speech conditioned on attacker-chosen text $y_A$, but the ASR
transcript is scored against a different, unpaired text $y_B$, with
$y_A\ne y_B$. This row uses a newly generated set of utterances rather than
directly rescoring the Standard audio; the model is not retrained. Its identity
and naturalness scores describe that output set, not an effect of changing
the scoring reference. Its WER measures mismatch with $y_B$, not adherence
error relative to the generation condition $y_A$.

Removing leakage reduces CAM similarity from 0.78 to 0.25, whereas retaining
leakage without text yields 0.57. Mismatched text yields CAM and ECAPA
similarities of 0.75 and 0.71 and UTMOS of 3.08, but WER against the unpaired
text is 99.17\%, compared with 3.80\% under the standard protocol. This high
error rate reflects mismatch with the scoring text and cannot be interpreted
as failure to follow the generation condition.

\begin{table}[t]
\caption{Leakage-conditioning and target-text controls for \stia speech generation using the Moshi frontend.}
\label{tab:conditioning-ablation}
\centering
\footnotesize
\setlength{\tabcolsep}{1.5pt}
\renewcommand{\arraystretch}{1.15}
\begin{tabular*}{\columnwidth}{@{\extracolsep{\fill}}lcccc@{}}
\toprule
Control & CAM $\uparrow$ & ECAPA $\uparrow$ & WER (\%) $\downarrow$ & UTMOS $\uparrow$ \\
\midrule
\textbf{Standard} & \textbf{0.78} & \textbf{0.79} & \textbf{3.80} & 2.92 \\
\addlinespace[3pt]
\quad - No-leak & \meandelta{0.25}{\downarrow}{0.54} & \meandelta{0.25}{\downarrow}{0.54} & \meandelta{12.94}{\uparrow}{9.14} & \meandelta{2.47}{\downarrow}{0.44} \\
\addlinespace[3pt]
\quad - No-text & \meandelta{0.57}{\downarrow}{0.21} & \meandelta{0.48}{\downarrow}{0.31} & --- & \meandelta{2.49}{\downarrow}{0.43} \\
\addlinespace[3pt]
\quad - Neither & \meandelta{0.22}{\downarrow}{0.57} & \meandelta{0.26}{\downarrow}{0.53} & --- & \meandelta{2.49}{\downarrow}{0.43} \\
\addlinespace[3pt]
\quad - Wrong text & \meandelta{0.75}{\downarrow}{0.04} & \meandelta{0.71}{\downarrow}{0.08} & \meandelta{99.17}{\uparrow}{95.37} & \meandelta{\mathbf{3.08}}{\uparrow}{0.16} \\
\bottomrule
\end{tabular*}
\end{table}

\subsection{Training Settings and Cost}
\label{sec:hyperparameters-runtime}

\begin{table}[t]
\caption{Training and sampling configuration of \stia and main-model training cost on the Moshi frontend.}
\label{tab:hyperparameters}
\centering
\footnotesize
\setlength{\tabcolsep}{3pt}
\begin{tabular}{p{1.42in}p{1.68in}}
\toprule
\textbf{Setting} & \textbf{Value} \\
\midrule
Optimizer / precision & AdamW / BF16 \\
Effective batch size & 64 \\
Training steps & 150,000 \\
Peak / final learning rate & $2.00\times10^{-4}$ / $2.00\times10^{-5}$ \\
Warmup / schedule & 1,000 steps / cosine decay \\
Main leak / model & 3.00 s / XL (0.60B) \\
Euler sampling steps & 50 \\
Guidance scale & 2.00 \\
\midrule
Training GPUs & 2 $\times$ A100-40GB \\
Training time & 110 h \\
\bottomrule
\end{tabular}
\end{table}

Table~\ref{tab:hyperparameters} summarizes the main training and sampling
settings and the training cost of \stia against Moshi.  The main configuration
uses a 3-second leak and the XL model with approximately 0.60B parameters.
All \stia training configurations use 150,000 steps. The main model targeting
Moshi was trained for 110 hours in BF16 on two NVIDIA A100 GPUs, each with
40 GB of memory. This time excludes preprocessing of the cached mT5 text
representations and Higgs speech latents.

\section{Discussion}
\label{sec:discussion}

\subsection{Security Implications}

SpInv established that speaker identity information can be recovered from
speech-language-model frontend representations~\cite{lu2026spinv}. Our results
show a further consequence: vocal cues retained in fragmented representations
can be reused to generate new, attacker-chosen content. Evaluation on three
representative frontends shows that a representation leak can expose identity
information and also provide a condition for generating speech with the
victim's vocal characteristics. A speaker may therefore face voice
impersonation risks even when the attacker cannot obtain the original recording.

Assessing voice impersonation risks requires examining both whose voice the
generated speech resembles and what it says. Table~\ref{tab:main-results} shows higher mean
speaker similarity and lower mean target-text WER for \stia than for the
SpInv+TTS baseline method, while mean UTMOS scores remain similar. The principal
observed gains are therefore in vocal-identity preservation and content
accuracy, with no large improvement in predicted naturalness. Privacy
evaluation should consider all three aspects together: high identity similarity
alone does not establish accurate delivery of the chosen text, and accurate
text does not establish resemblance to the victim's voice.

\subsection{Defensive Directions}

Because the attack assumes read access to frontend representations, limiting
that access is a direct defensive measure. Authenticated encryption can protect
representations in transit, while process isolation, restricted read
permissions, and short retention can reduce exposure at endpoints. Protection
must also account for services or relays that legitimately receive the
representations, since transport encryption alone cannot prevent their reuse
by a recipient. Systems should therefore identify which components can read or
store representations and restrict access accordingly. Shorter exposure can
reduce the information available, but Sec.~\ref{sec:leakage-ablation} shows that
even the shortest tested fragments retain useful speaker cues.

A complementary direction is to alter speaker identity cues before frontend
encoding. In the Moshi experiment, McAdams anonymization lowers both mean
speaker-similarity scores for the fixed attacker, along with the UTMOS of the
generated speech (Table~\ref{tab:defense}). This shows that the transformation
can weaken the evaluated attacker, but does not establish that identity
information has been removed. Because the attacker was trained only on
unanonymized speech, the decrease may also partly reflect differences between
its inputs and training data. Further defense evaluation should therefore
include attackers retrained on anonymized inputs and measure the impact on
legitimate backend tasks; neither aspect is evaluated here.

In applications that use voice for authentication, an additional authentication
factor or independent confirmation tied to a specific transaction could help
limit the consequences of forged speech. Such measures address the
application-level impact of voice impersonation and complement protection of
the frontend representations themselves.

\subsection{Limitations}

\noindent\textbf{Evaluation coverage.}
The experiments use speaker-disjoint subsets of Emilia and attack models
trained for each target frontend. Three representative frontends are evaluated
in English, with a separate Chinese evaluation on Moshi. The results therefore
establish generalization to unseen speakers within the tested
settings; generalization across datasets or languages, and transfer across
frontends without retraining, remain untested.

\noindent\textbf{Output-codec dependence.}
Although \stia accommodates multiple frontend representation formats, speech
generation uses the frozen Higgs codec's continuous pre-quantization latent
space. The vocal characteristics and quality of the generated speech therefore
also depend on what the Higgs codec can represent and reconstruct. Alternative
output codecs have not been evaluated.

\noindent\textbf{Measurement limitations.}
Speaker similarity, ASR error rates, and UTMOS provide automatic measurements
of vocal identity, text accuracy, and naturalness, but may not fully agree with
listener judgments. The reported similarities are not authentication success
rates, and UTMOS predicts naturalness rather than collecting listener ratings.
Listener studies and evaluation against operational authentication systems are
needed to determine practical impersonation effectiveness.

\section{Ethics Considerations}
\label{sec:ethics}

This study uses public Emilia data for offline privacy measurement; the
dataset is distributed under CC BY-NC 4.0. We synthesize speech resembling
dataset speakers without recruiting participants, collecting private recordings,
or constructing real-world identity profiles. Institutional consultation
confirmed that formal ethics approval was not required for this study.
Experiments remain within controlled, dataset-based settings and do not deploy
attacks against external services.

Public availability does not imply speaker consent to voice cloning.
Representation inversion could enable unauthorized impersonation or fraud.
To limit individual exposure through experimental results, this paper reports
aggregate metrics rather than identifiable voice demonstrations. Future releases
should respect dataset licenses and restrict distribution of identifiable audio
and reusable attack models. The study aims to expose these risks and inform
more privacy-preserving speech interfaces.

\section{Conclusion}
\label{sec:conclusion}

We introduced \stia for arbitrary-text speaker impersonation from leaked
frontend representation fragments of end-to-end speech language models.
The method combines cross-utterance training with conditional flow matching
to reuse vocal cues in representations without the victim's waveform or source
transcript. Experiments on three representative frontends show higher mean speaker similarity and lower mean target-text
WER than the SpInv+TTS baseline method, exposing voice impersonation risks
from representation leakage.

Future work should pursue three directions. First, extend evaluation to
additional datasets, languages, and realistic deployment conditions. Second,
assess practical impersonation effectiveness through listener studies and
tests against operational authentication systems. Third, investigate stronger
voice anonymization and frontend privacy protections, jointly evaluating their
resistance to adaptive attackers and their impact on legitimate speech tasks.

\bibliographystyle{IEEEtran}
\bibliography{references}

\end{document}